# Processing-Controlled Structural Uniformity and Oxide-Ion Conduction in $Na_{0.52}Bi_{0.47}TiO_3$ Ceramics Probed by $Eu^{3+}$ Photoluminescence

Zhouyang He[1], Xiaoou Sun[1], Xinyue Wang[1], Duanting Yan[1*]

[1] State Key Laboratory of Integrated Optoelectronics, Key Laboratory of UV Light-Emitting Materials and Technology of Ministry of Education, School of Physics, Northeast Normal University

Changchun, China

*Corresponding authors.

Email: *yandt677@nenu.edu.cn

## Abstract

Sodium bismuth titanate (NBT) is a promising oxide-ion conductor, but its electrical conductivity is highly sensitive to small changes in A-site stoichiometry and processing history. This sensitivity can reduce sample-to-sample reproducibility. Here we examine how precursor mixing controls structural uniformity and ionic transport in $Na_{0.52}Bi_{0.47}TiO_3$ ceramics. Dry grinding, wet grinding with ethanol, and ball milling were compared by X-ray diffraction, electron microscopy, energy-dispersive spectroscopy, $Eu^{3+}$ photoluminescence excitation spectroscopy, and electrochemical impedance spectroscopy. All processed powders and ceramics form the perovskite NBT phase within the detection limit of XRD. However, the microstructure, surface A-site cation ratio, $Eu^{3+}$ excitation spectra, and electrical response change strongly with the mixing route. Continuous monitoring of $Eu^{3+}$ excitation spectra at different emission wavelengths reveals different distributions of local $Eu^{3+}$ environments. Larger spectral-shape variations are consistent with lower structural uniformity and stronger local distortion. Dry-ground samples show higher bulk conductivity than wet-ground samples, whereas wet-ground samples show much lower grain-boundary resistance. At 600 ℃, the dry-60 min sample reaches a bulk conductivity of 13.54 mS $cm^{-1}$, while wet-30 min shows the highest grain-boundary conductivity of 13.72 mS $cm^{-1}$. These results suggest a processing-driven trade-off between bulk defect generation and grain-boundary blocking. Based on this processing understanding, Ca was introduced at the A site in $Na_{0.52}Bi_{0.47-x}Ca_xTiO_3$. The x = 0.04 sample reaches 8.35 mS $cm^{-1}$ at 500 ℃ and 18.98 mS $cm^{-1}$ at 600 ℃.

**Keywords:** sodium bismuth titanate; oxide-ion conductor; structural uniformity; $Eu^{3+}$; photoluminescence

## 1. Introduction

Oxide-ion conductors are key functional materials for solid oxide fuel cells, oxygen sensors, gas sensors, memristive devices, and oxygen separation membranes[1-6]. Among them, sodium bismuth titanate, $Na_{0.5}Bi_{0.5}TiO_3$ (NBT), has attracted increasing attention because it is lead-free, chemically stable, and can exhibit high oxide-ion mobility at intermediate temperatures[7-9]. Although NBT was first developed mainly as a ferroelectric and piezoelectric ceramic, its high leakage current has been linked to oxygen-ion migration rather than electronic conduction in suitable compositions[10]. The conductivity of NBT-based materials is unusually sensitive to the A-site Na/Bi ratio. Bi-rich compositions tend to be insulating, whereas Na-rich or Bi-deficient compositions can become oxide-ion conductors[10-12]. Atomistic simulations and defect-chemistry studies further show that small A-site nonstoichiometry changes can alter both carrier concentration and migration behavior [11-13]. Several doping strategies have therefore been developed. A-site dopants such as Ca, Sr, and $Sm^{3+}$, and B-site dopants such as Mg, $Sc^{3+}$, and $Al^{3+}$, have been used to tune oxygen-vacancy concentration and improve conductivity [14-19].

Most previous work emphasizes nominal composition and dopant concentration. In contrast, processing-induced structural uniformity receives less attention. This omission is important for NBT. Both Na- and Bi-containing precursors can undergo compositional redistribution or volatilization during heat treatment. Local variation in A-site chemistry may generate nonuniform oxygen-vacancy distributions, local lattice distortion, and heterogeneous grain boundaries. These effects can change both bulk and grain-boundary transport even when the nominal composition is fixed.

Conventional XRD and averaged composition analysis provide useful phase and composition information, but they are not sensitive enough to resolve subtle local heterogeneity in a multi-cation perovskite. $Eu^{3+}$ photoluminescence offers a complementary probe because the $Eu^{3+}$ charge-transfer band and 4f-4f transitions respond to local coordination, covalency, and crystal-field symmetry. By monitoring photoluminescence excitation (PLE) spectra at a series of emission wavelengths, one can detect whether different $Eu^{3+}$ emitting environments have similar or different excitation responses. A larger wavelength-dependent spectral variation indicates a broader distribution of local environments. This method does not directly quantify oxygen-vacancy concentration, but it can sensitively track local structural heterogeneity.

In this work, we use $Na_{0.52}Bi_{0.47}TiO_3$ as a model composition to examine how precursor mixing routes affect structural uniformity and oxide-ion conduction. Dry grinding, wet grinding with ethanol, and ball milling are compared. We correlate XRD, SEM, EDS, $Eu^{3+}$ PLE spectra, and impedance spectroscopy. We then use the processing insight to prepare Ca-doped $Na_{0.52}Bi_{0.47-x}Ca_xTiO_3$ ceramics with improved bulk conductivity. The central goal is to clarify why nominally similar NBT ceramics can show different electrical responses and how processing can be used to control this variability.

## 2. Experimental Section

### 2.1 Sample preparation

$Bi_2O_3$, rutile $TiO_2$, and anhydrous $Na_2CO_3$ were used as starting reagents. Before weighing, $Bi_2O_3$ and $Na_2CO_3$ were dried at 300 °C for 5 h, and $TiO_2$ was dried at 800 °C for 2 h. The nominal composition for the processing study was $Na_{0.52}Bi_{0.47}TiO_3$. The weighed powders were mixed by two routes. In the dry route, the powders were ground without ethanol for 60 or 90 min, and the samples are denoted dry-60 min and dry-90 min. In the wet route, ethanol was used as the grinding medium for 30 or 60 min, and the samples are denoted wet-30 min and wet-60 min. A ball-milled wet-ground sample is denoted

wet-60 min-ball. The mixed powders were calcined at 800 °C for 2 h. The ceramics were sintered at 1050 °C for 2 h. For the Ca-doped series, the nominal composition was $Na_{0.52}Bi_{0.47-x}Ca_xTiO_3$ (x = 0.00, 0.01, 0.02, 0.03, 0.04, and 0.05). The supplied draft does not specify the Ca precursor, $Eu^{3+}$ probe concentration, pellet pressing condition, or whether the same mixing route was used for all Ca-doped samples.

### 2.2 Characterization

Phase formation was examined by X-ray diffraction (XRD). Powder and ceramic microstructures were observed by scanning electron microscopy (SEM). Elemental distributions and surface compositions were analyzed by energy-dispersive spectroscopy (EDS). $Eu^{3+}$ PLE spectra were collected by monitoring the $Eu^{3+}$ electric-dipole emission in the 609-615 nm range. The relative intensity ratio between the $O^{2-}$-$Eu^{3+}$ charge-transfer band and a 4f-4f excitation band is defined as $R_1 = I_{CT}/I_{4f}$. Impedance spectra were measured to separate bulk and grain-boundary responses. Conductivities were analyzed using Arrhenius plots.

## 3. Results and Discussion

### 3.1 Phase formation and processing-dependent microstructure

Figure 1 compares the XRD patterns of powders and ceramics prepared by different mixing routes. All samples show the characteristic diffraction peaks of rhombohedral NBT, and no secondary phase is detected within the resolution of XRD. This result indicates that the selected calcination and sintering conditions are sufficient to form the main perovskite phase for all processing routes. The ceramic peaks shift slightly to higher 2θ values relative to the powder peaks. This shift indicates a small decrease in lattice spacing after sintering. However, XRD alone cannot identify whether the shift originates from Na/Bi redistribution, volatilization, oxygen-vacancy formation, residual strain, or a combination of these factors. Therefore, this shift is treated as evidence for a subtle lattice change rather than direct proof of a specific defect process.

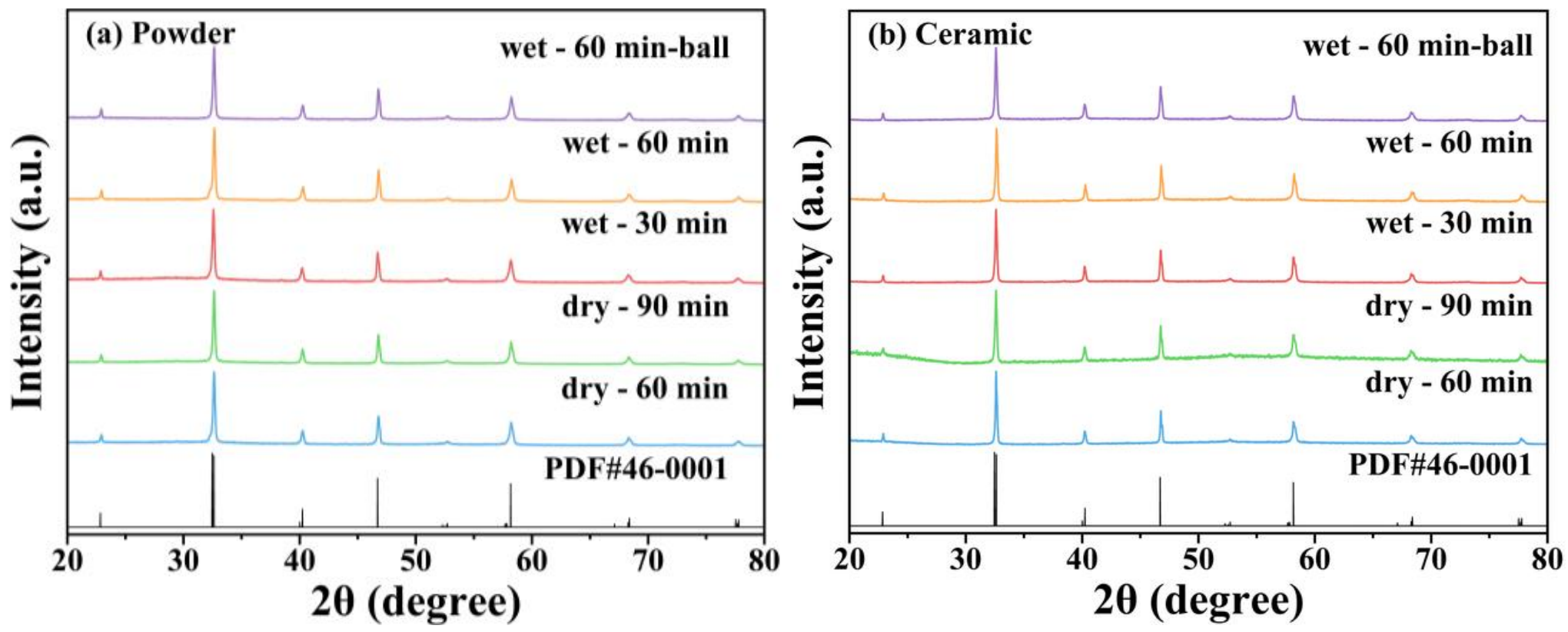

**Figure 1.** XRD patterns of $Na_{0.52}Bi_{0.47}TiO_3$ prepared by different precursor mixing routes: (a) calcined powders and (b) sintered ceramics.

The powder morphology depends strongly on the mixing route (Figure 2). The dry-60 min and dry-90 min powders contain relatively separated particles and show less obvious agglomeration. In contrast, the wet-ground powders contain larger agglomerates and less uniform particle clusters. The wet-60 min-ball powder contains smaller particles than the hand-ground wet samples, which is consistent with the stronger mechanical fragmentation produced by ball milling. These observations suggest that ethanol-assisted wet grinding does not simply improve dispersion in this system. Residual solvent, drying behavior, or capillary aggregation may promote powder agglomeration before calcination.

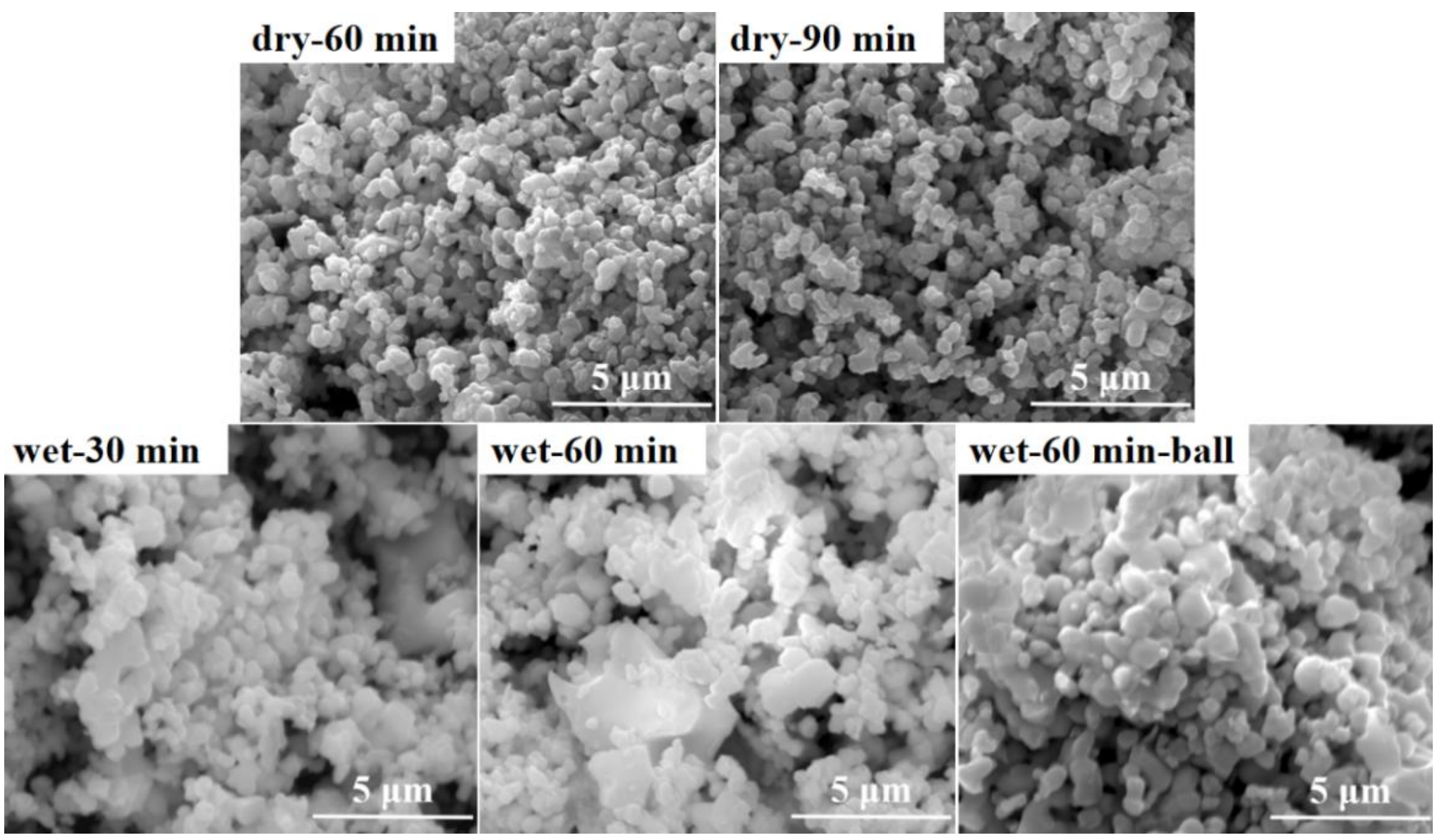

**Figure 2.** SEM images of $Na_{0.52}Bi_{0.47}TiO_3$ powders prepared by dry grinding, wet grinding, and wet ball milling.

EDS mapping of representative powders further shows that the surface cation distribution changes with processing (Figure 3). The wet-30 min powder contains less Na-rich surface signal than the dry-60 min powder in the mapped region. The corresponding EDS spectra (Figure 4) give different semi-quantitative A-site ratios. The wet-30 min powder has nearly comparable Na and Bi contents in the analyzed surface region, whereas the dry-60 min powder is more Na-rich relative to Bi. These results show that dry and wet processing lead to different surface A-site compositions after calcination.

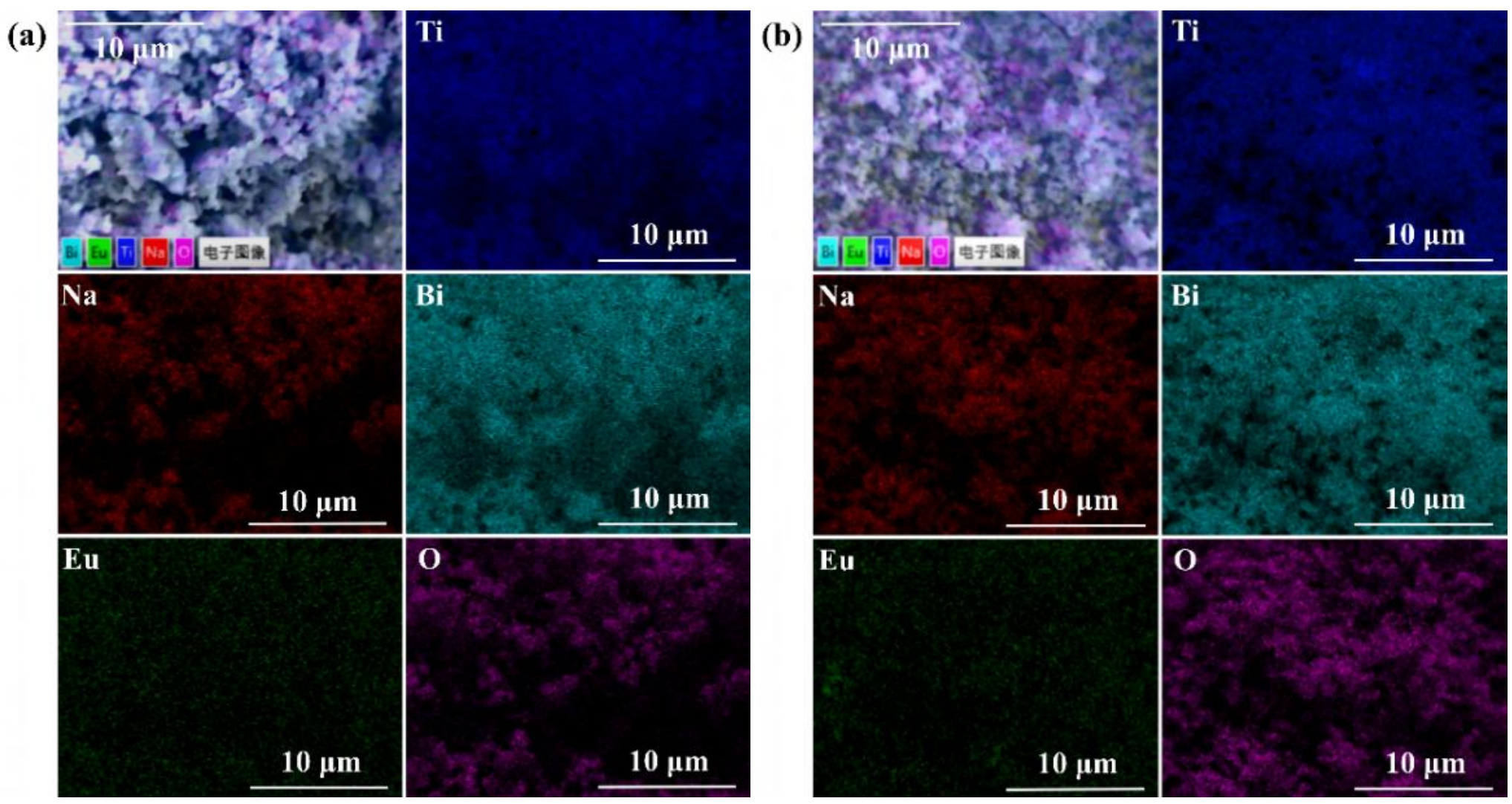

**Figure 3.** EDS elemental maps of representative $Na_{0.52}Bi_{0.47}TiO_3$ powders: (a) wet-30 min and (b) dry-60 min.

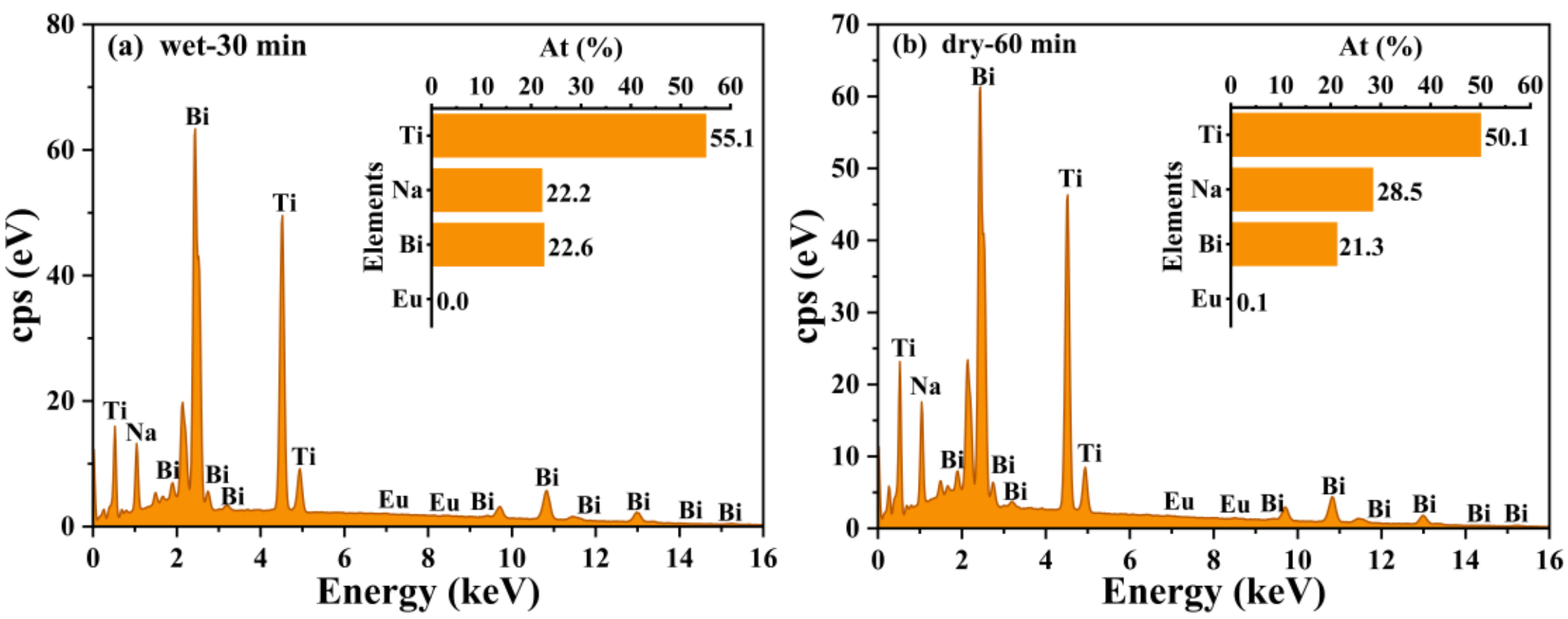

**Figure 4.** EDS spectra and semi-quantitative elemental contents of representative $Na_{0.52}Bi_{0.47}TiO_3$ powders: (a) wet-30 min and (b) dry-60 min.

The ceramic microstructures inherit part of the powder-processing history (Figure 5). Dry-ground ceramics show larger average grain size and more similar grain morphology than the wet-ground ceramics. The wet-ground ceramics show broader grain-size distributions, with large grains distributed among smaller grains. The wet-60 min-ball ceramic contains the finest overall microstructure among the compared samples. These differences indicate that the precursor mixing route influences sintering and grain growth. The effect may arise from differences in particle contact, agglomeration, A-site surface composition, and local defect chemistry.

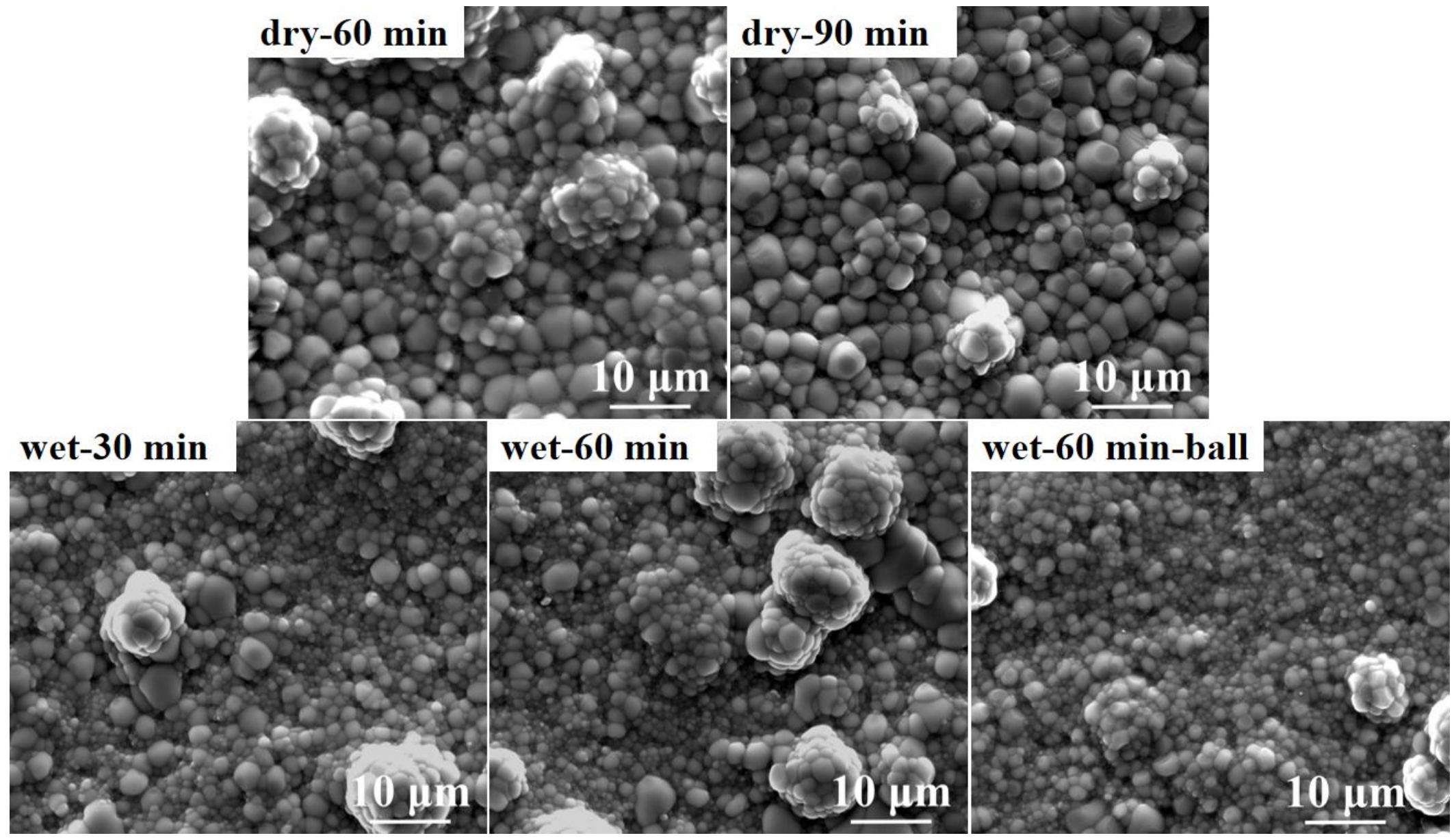

**Figure 5.** SEM images of $Na_{0.52}Bi_{0.47}TiO_3$ ceramics prepared by different mixing routes.

EDS mapping of ceramics also indicates processing-dependent A-site distributions (Figure 6). The wet-30 min ceramic shows larger local contrast between Na and Bi signals in the selected region. The dry-60 min ceramic shows a more uniform Na/Bi distribution over the mapped area. The EDS spectra of all ceramics (Figure 7) show that the surface Na/Bi ratio varies with grinding route and time. The values reported from the EDS insets are approximately 1.13, 1.12, 0.85, 1.08, and 1.05 for dry-60 min, dry-90 min, wet-30 min, wet-60 min, and wet-60 min-ball, respectively.

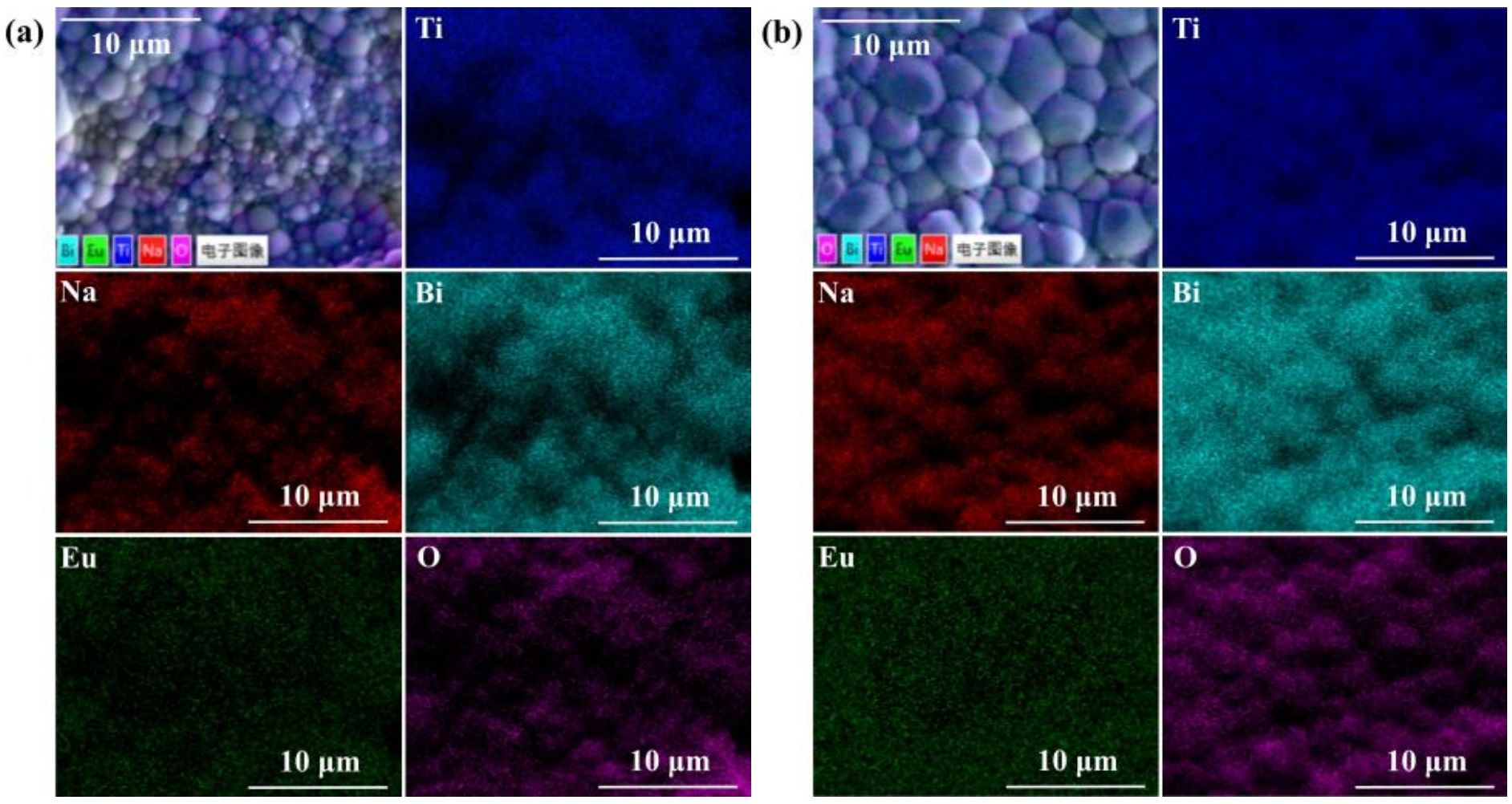

**Figure 6.** EDS elemental maps of representative $Na_{0.52}Bi_{0.47}TiO_3$ ceramics: (a) wet-30 min and (b) dry-60 min.

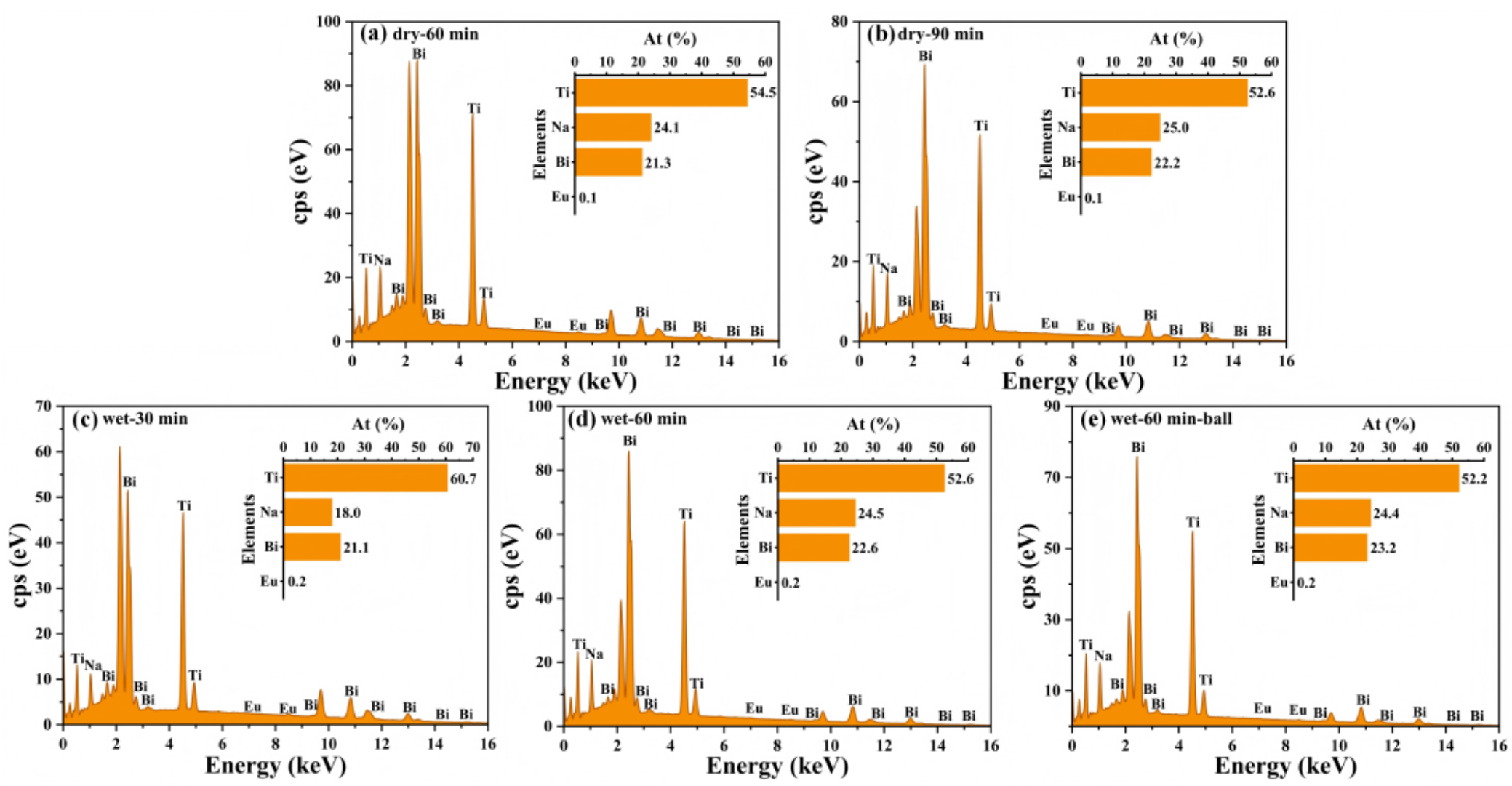

**Figure 7.** EDS spectra and semi-quantitative elemental contents of $Na_{0.52}Bi_{0.47}TiO_3$ ceramics prepared by different mixing routes.

## 3.2 $Eu^{3+}$ PLE response as a probe of structural uniformity

$Eu^{3+}$ PLE spectra provide a sensitive view of local structural heterogeneity that is not resolved by XRD. In these spectra, the broad charge-transfer (CT) band reflects $O^{2-}$-$Eu^{3+}$ interactions and is sensitive to local coordination and covalency. The sharp 4f-4f excitation lines are mainly intra-ionic transitions, but their relative intensities and line shapes can still respond to local symmetry. When the PLE spectrum changes little with the monitored emission wavelength, the $Eu^{3+}$ emitting centers experience similar local environments. When the PLE spectrum changes strongly with monitoring wavelength, the sample contains a broader distribution of $Eu^{3+}$ local environments. Therefore, the wavelength-dependent change in $R_1 = I_{CT}/I_{4f}$ is used here as a qualitative indicator of structural uniformity.

Figure 8 shows normalized PLE spectra of powders monitored from 609 to 615 nm. At the same grinding time, dry-60 min and wet-60 min powders show clearly different CT-band shapes and different $R_1$ variations. The dry-ground powders show larger spectral-shape changes than the wet-ground powders, which suggests a broader distribution of local $Eu^{3+}$ environments. Extending dry grinding from 60 to 90 min produces a similar overall spectral trend but a stronger $R_1$ variation. In the wet-ground series, longer grinding and ball milling also increase the spectral-shape variation. These observations are consistent with the microstructural evidence: stronger mechanical dispersion or smaller particle size can enhance local compositional redistribution during calcination, while agglomeration can partly suppress it.

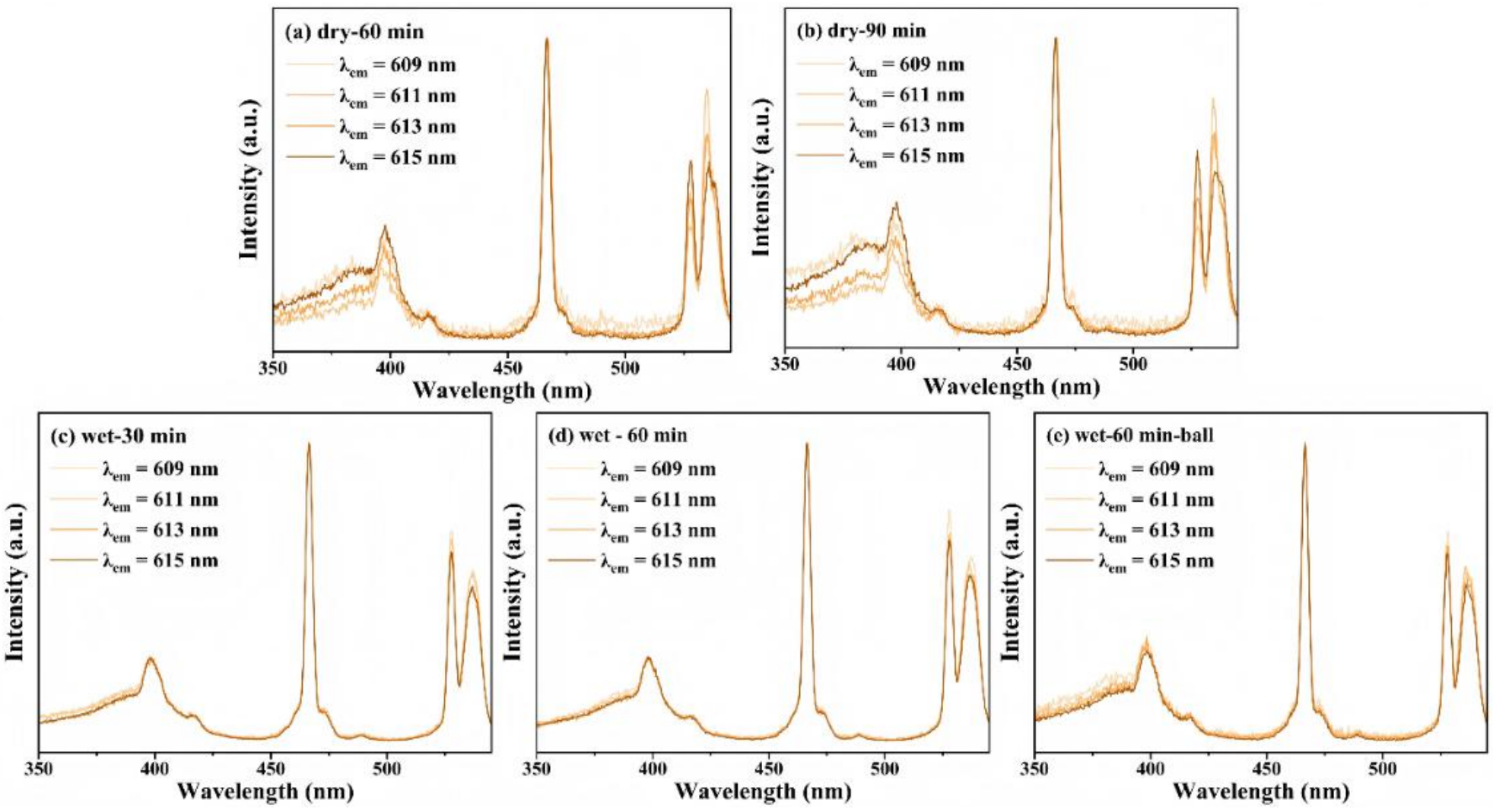

**Figure 8.** Normalized PLE spectra of $Eu^{3+}$-probed $Na_{0.52}Bi_{0.47}TiO_3$ powders prepared by different mixing routes.

The ceramics show similar processing-dependent PLE behavior (Figure 9). The dry-60 min ceramic shows stronger spectral variation than the wet-30 min ceramic, indicating lower structural uniformity. Among the dry-ground ceramics, dry-60 min shows stronger variation than dry-90 min. Among the wet-ground ceramics, the spectral variation decreases in the order wet-60 min-ball, wet-60 min, and wet-30 min. Thus, the powder-processing route continues to affect the local structure after sintering. The PLE data suggest that structural uniformity in NBT is not determined only by the final sintering temperature. It also depends on the precursor state established before calcination and sintering.

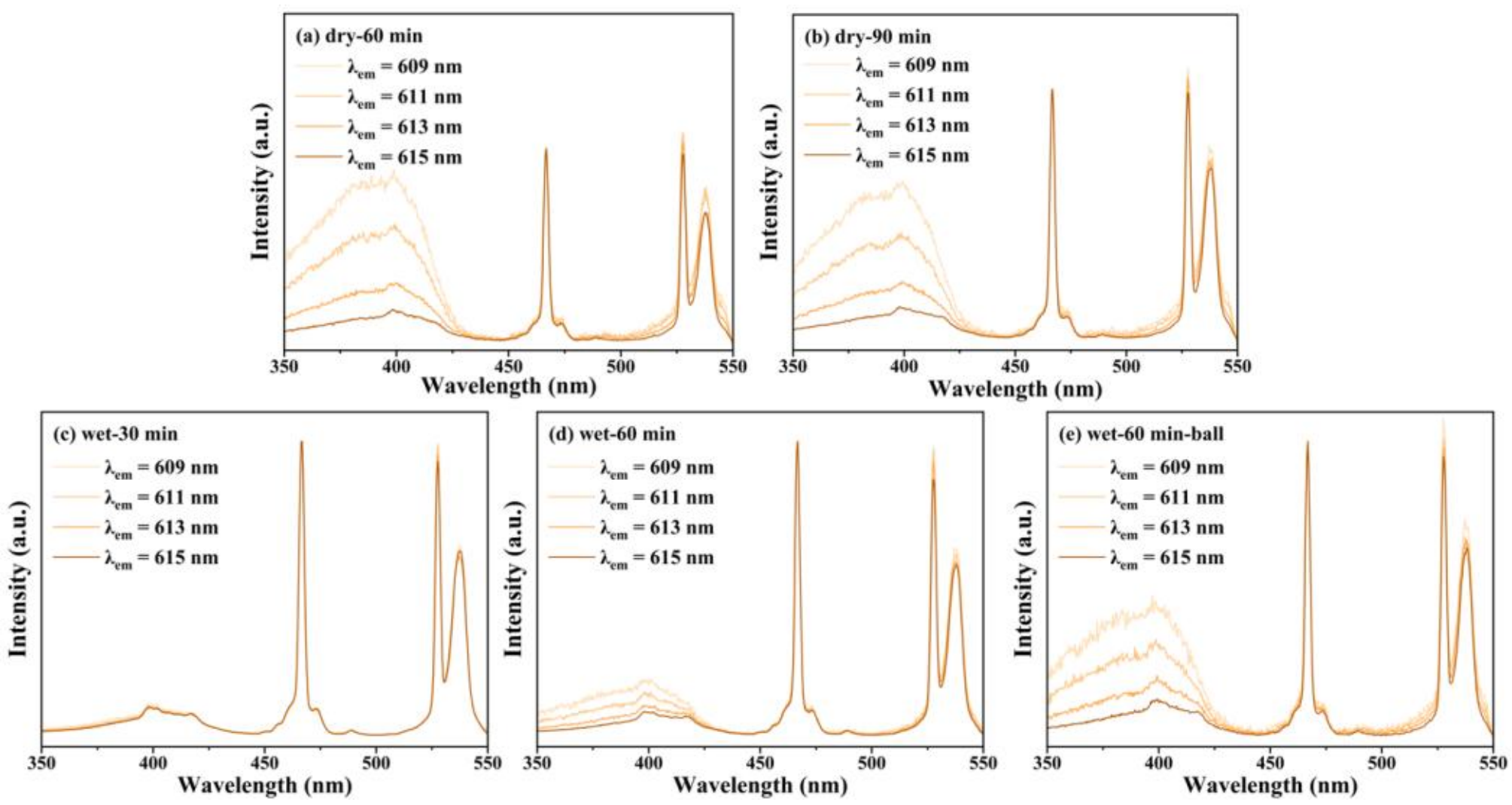

**Figure 9.** Normalized PLE spectra of $Eu^{3+}$-probed $Na_{0.52}Bi_{0.47}TiO_3$ ceramics prepared by different mixing routes.

### 3.3 Correlation between structural uniformity and electrical transport

Impedance spectroscopy reveals that the processing route affects bulk and grain-boundary transport in different ways. Figure 10 shows representative impedance spectra measured at 400 °C. The high-frequency response corresponds to the bulk contribution, whereas the lower-frequency arc is assigned to the grain-boundary contribution. The dry-ground samples have smaller bulk resistance than the wet-ground samples. However, they also show larger grain-boundary resistance. The wet-ground samples have slightly higher bulk resistance but much lower grain-boundary resistance. This contrast shows that a processing route favorable for bulk conduction may be unfavorable for grain-boundary conduction.

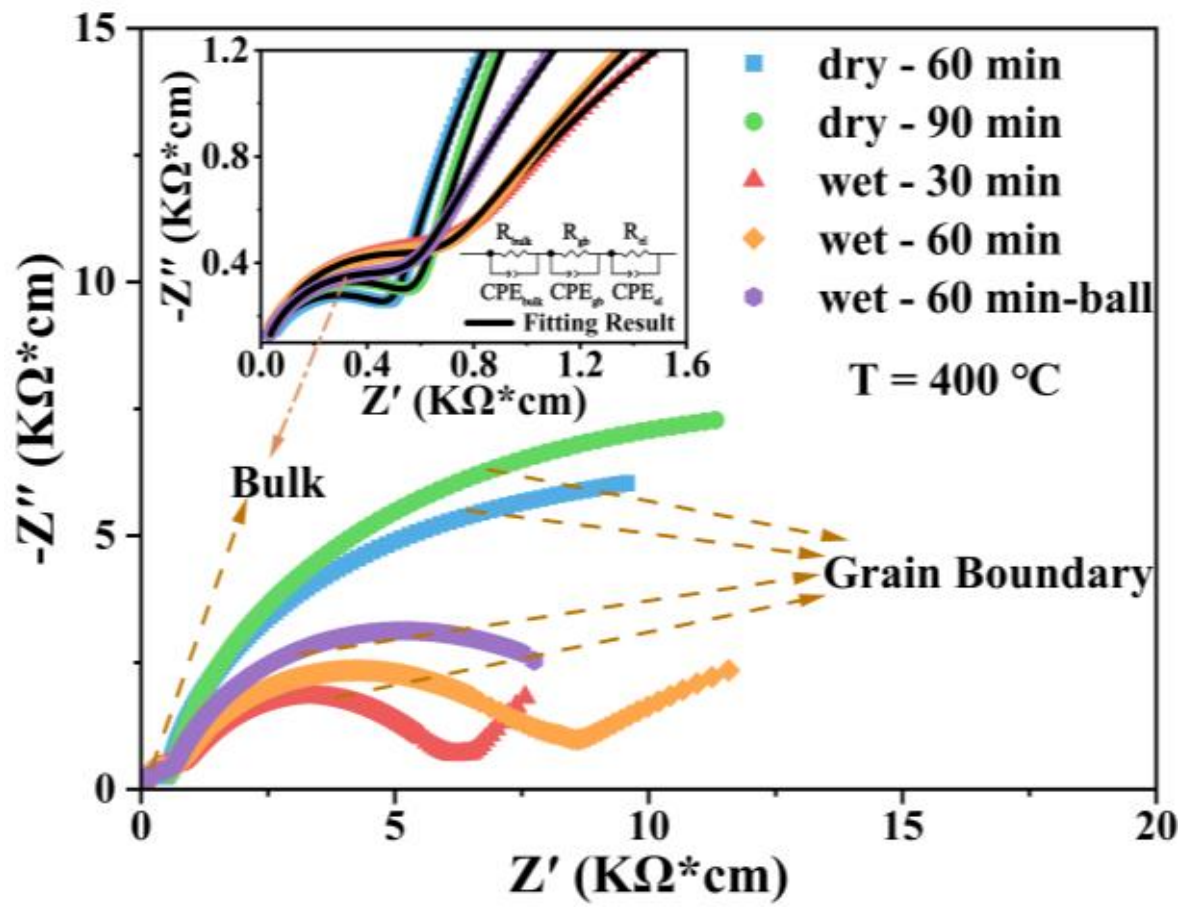

**Figure 10.** Impedance spectra of $Na_{0.52}Bi_{0.47}TiO_3$ ceramics prepared by different mixing routes at 400 °C.

Arrhenius plots of the bulk and grain-boundary conductivities are shown in Figure 11. Above about 400 °C, the bulk conductivity difference between dry- and wet-ground samples becomes more pronounced. At 600 °C, dry-60 min shows the highest bulk conductivity of 13.54 mS $cm^{-1}$, while wet-30 min shows the lowest value of 6.7 mS $cm^{-1}$. The overall bulk-conductivity order is dry-60 min > dry-90 min > wet-60 min-ball > wet-60 min > wet-30 min. This order agrees qualitatively with the PLE result that stronger local heterogeneity is associated with higher bulk conductivity. A plausible explanation is that enhanced A-site compositional redistribution produces more charge-compensating oxygen vacancies or lowers the effective migration barrier in the bulk.

The grain-boundary trend is almost opposite. At 600 °C, wet-30 min shows the highest grain-boundary conductivity of 13.72 mS $cm^{-1}$, whereas dry-90 min shows the lowest value of 0.56 mS $cm^{-1}$. Within the wet-ground series, the grain-boundary resistance increases in the order wet-30 min, wet-60 min, and wet-60 min-ball. This trend suggests that stronger local heterogeneity or stronger A-site redistribution can increase bulk conductivity but may also create more blocking grain boundaries. Possible causes include A-site segregation, nonuniform oxygen-vacancy distribution near interfaces, or changes in grain size and grain-boundary area. The data therefore support a processing-driven trade-off between bulk carrier generation and grain-boundary transport.

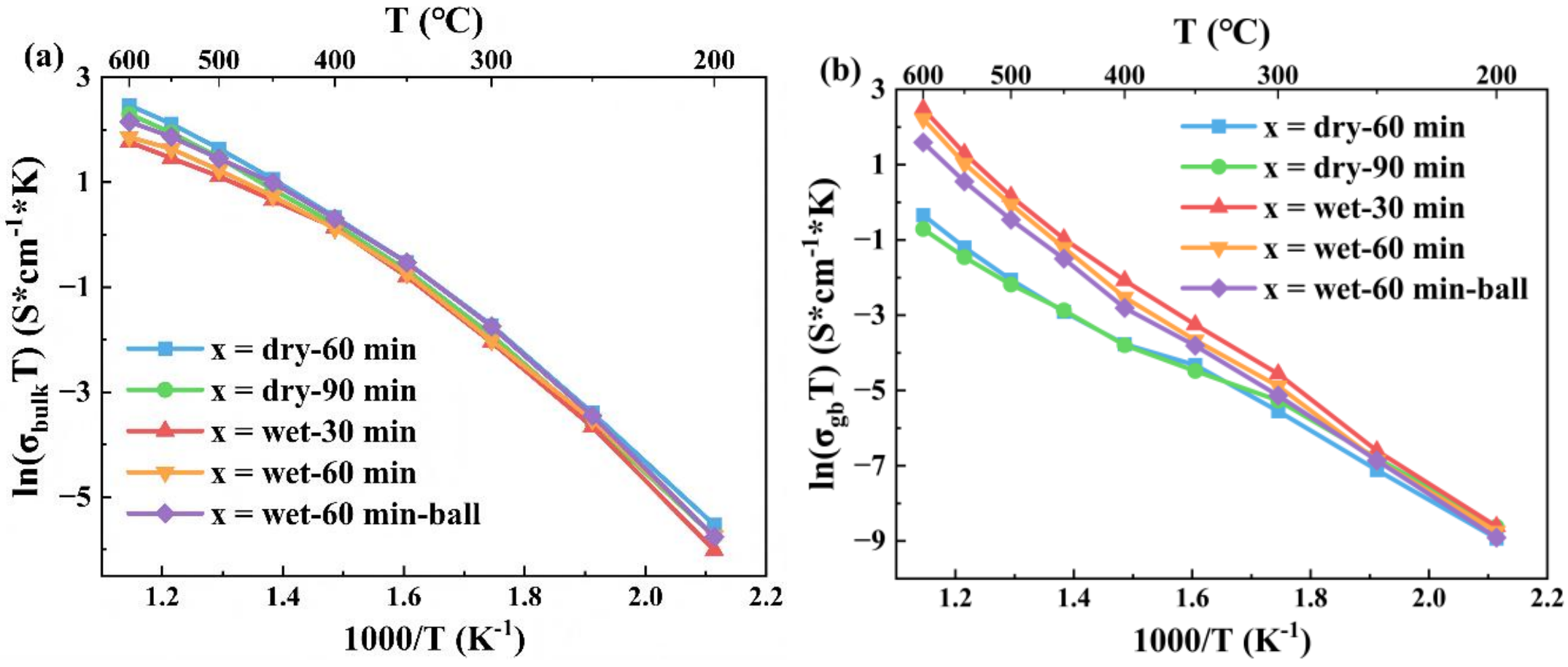

**Figure 11.** Arrhenius plots of (a) bulk conductivity and (b) grain-boundary conductivity for $Na_{0.52}Bi_{0.47}TiO_3$ ceramics prepared by different mixing routes.

### 3.4 $Ca^{2+}$ acceptor doping based on processing control

The processing study indicates that bulk conductivity can be enhanced when the A-site defect chemistry and local structure are tuned. $Ca^{2+}$ was therefore introduced at the A site in $Na_{0.52}Bi_{0.47-x}Ca_xTiO_3$. $Ca^{2+}$ substitution for $Bi^{3+}$ is expected to act as acceptor doping and to promote charge-compensating oxygen vacancies. Figure 12a shows that Ca doping improves the bulk conductivity relative to the undoped composition, with the highest conductivity at x = 0.04. Further increasing the Ca content to x = 0.05 does not provide additional improvement, which suggests that an optimum exists between vacancy generation and defect association or local lattice distortion.

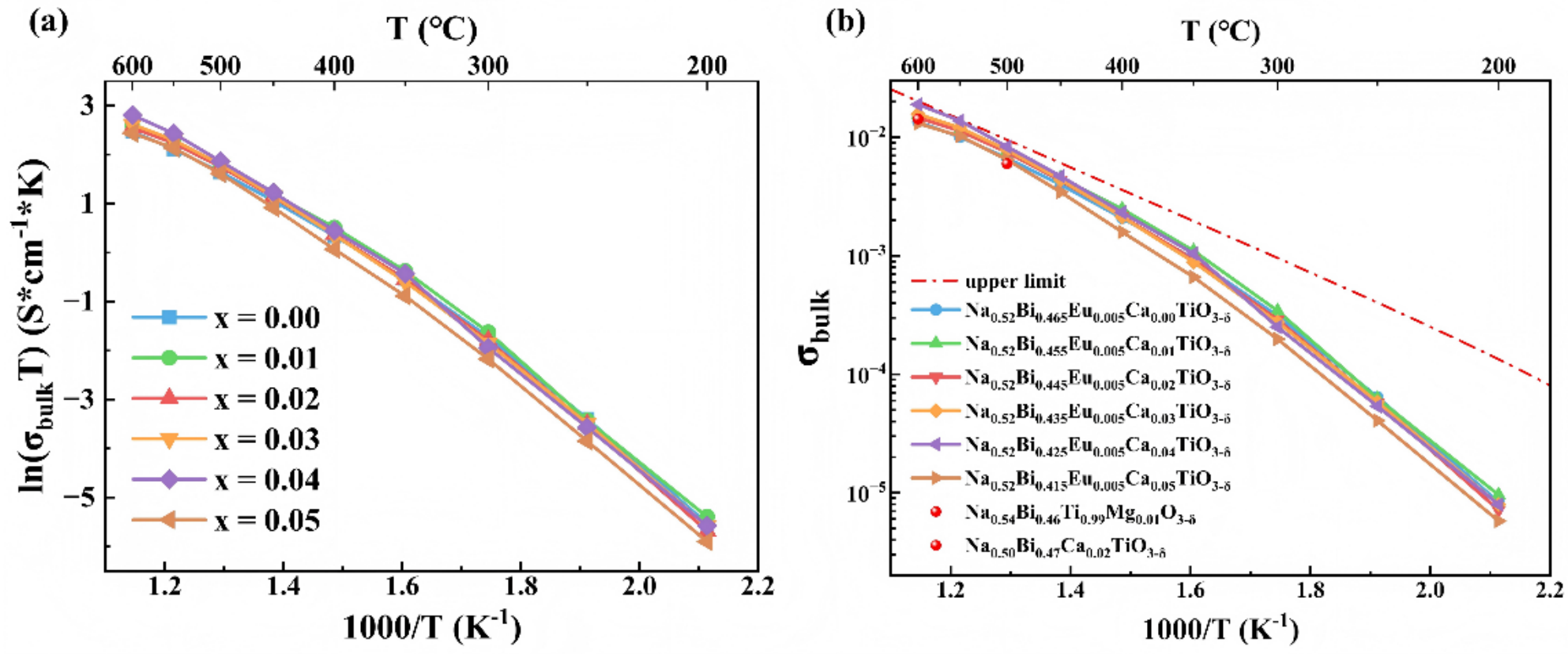

**Figure 12.** Bulk conductivity of $Na_{0.52}Bi_{0.47-x}Ca_xTiO_3$ ceramics: (a) Arrhenius plots for x = 0.00-0.05 and (b) comparison with reported literature benchmarks.

Table 1 summarizes the apparent activation energies in two temperature ranges. In the 200-350 °C range, the activation energy remains nearly constant at 0.85-0.88 eV. In the 350-600 °C range, it increases from 0.57 eV for x = 0.00 to 0.64 eV for x = 0.05. If the high-temperature value mainly reflects migration and the low-temperature value contains both migration and association contributions, the difference between the two values decreases with Ca content. This trend is consistent with a reduced defect-association contribution at lower temperature.

**Table 1. Apparent activation energies for bulk conduction in $Na_{0.52}Bi_{0.47-x}Ca_xTiO_3$ ceramics.**

| Sample | $E_\alpha$ (200-350 °C) (eV) | $E_\alpha$ (350-600 °C) (eV) |
| --- | --- | --- |
| x = 0.00 | 0.85 | 0.57 |
| x = 0.01 | 0.86 | 0.56 |
| x = 0.02 | 0.88 | 0.59 |
| x = 0.03 | 0.85 | 0.61 |
| x = 0.04 | 0.87 | 0.62 |
| x = 0.05 | 0.85 | 0.64 |

The best sample, $Na_{0.52}Bi_{0.43}Ca_{0.04}TiO_3$, reaches 8.35 mS $cm^{-1}$ at 500 °C and 18.98 mS $cm^{-1}$ at 600 °C. These values are higher than the benchmark values cited in the draft, namely 6 mS $cm^{-1}$ at 500 °C and 14.3 mS $cm^{-1}$ at 600 °C [15,16]. The improvement is substantial but should be discussed in the context of processing route, grain-boundary resistance, and measurement normalization.

## 4. Conclusions

This work shows that precursor mixing strongly affects structural uniformity and oxide-ion conduction in $Na_{0.52}Bi_{0.47}TiO_3$ ceramics. XRD confirms phase-pure perovskite NBT for all processing routes, but SEM, EDS, $Eu^{3+}$ PLE spectra, and impedance spectroscopy reveal clear processing-dependent differences. Dry grinding produces higher bulk conductivity, with dry-60 min reaching 13.54 mS $cm^{-1}$ at 600 °C. Wet grinding lowers grain-boundary resistance, and wet-30 min reaches a grain-boundary conductivity of 13.72 mS $cm^{-1}$ at 600 °C. $Eu^{3+}$ PLE spectra show that samples with stronger wavelength-dependent spectral variation have broader distributions of local environments. This behavior is consistent with lower structural uniformity and stronger local defect heterogeneity. The results suggest that A-site cation redistribution during heat treatment can enhance bulk conduction but may also increase grain-boundary blocking. Based on this understanding, $Ca^{2+}$ acceptor doping was used to improve bulk conductivity. $Na_{0.52}Bi_{0.43}Ca_{0.04}TiO_3$ reaches 8.35 mS $cm^{-1}$ at 500 °C and 18.98 mS $cm^{-1}$ at 600 °C. These findings highlight structural uniformity as a key processing variable for reproducible high conductivity in NBT-based oxide-ion conductors.